\documentclass[aps,prl,twocolumn,superscriptaddress,floatfix,longbibliography]{revtex4-2}

\usepackage{amsmath,amssymb,bm}
\usepackage{graphicx}
\usepackage{booktabs}
\usepackage{xcolor}
\usepackage{hyperref}
\usepackage{url}
\definecolor{apslink}{RGB}{46,48,146}
\hypersetup{colorlinks=true, citecolor=apslink, linkcolor=apslink, urlcolor=apslink}
\graphicspath{{figures/}}

\newcommand{\qs}{q_{*}}
\newcommand{\xp}{x^\prime}

\newcommand{\dd}{\,\mathrm d}
\newcommand{\Prob}{\mathbb P}

\newcommand{\xP}{x_P}
\newcommand{\xF}{x_F}

\begin{document}

\title{Learning by Confusion Locates a Parametrization-Dependent Fisher
Information Maximum}

\author{George Issa}
\affiliation{Department of Physics and Astronomy, University of California, Davis, California 95616, USA}

\author{Richard Scalettar}
\affiliation{Department of Physics and Astronomy, University of California, Davis, California 95616, USA}

\begin{abstract}
Learning by confusion (LBC) is a machine-learning method for locating
phase transitions without assuming a known order parameter.
Its characteristic accuracy curve contains a central maximum that is
interpreted as the transition point, but what determines the location
and finite-size scaling of this maximum has remained unclear.
We derive an exact balance condition for the Bayes-optimal confusion
peak and show that, for equilibrium systems with a single swept
parameter, the peak closely follows the maximum of the corresponding
Fisher information.
This immediately gives its finite-size scaling and reveals an unexpected
dependence on how the data are sampled: the same configurations arranged
on grids uniform in $T$, ${\rm ln} T$, or $\beta=1/T$ can yield
different finite-size pseudocritical trajectories.
Tests on the Ising, Blume-Capel, Potts, and Baxter-Wu models confirm the
prediction across continuous, tricritical, and first-order transitions.
Our results identify the statistical quantity measured by LBC and show
that finite-size machine-learning estimates can depend on the
parametrization of the training data as well as on the underlying
physics.
\end{abstract}

\maketitle
\textit{Introduction.---} Classical phase transitions are associated
with non-analyticities of derivatives of the partition function
${\cal Z}$.
Since ${\cal Z}$ is just the sum of terms involving the exponential of
the energy of configurations, a system of finite linear size $L$, and
hence a finite number of terms in the sum, has no sharp phase
transition.
Nevertheless, finite systems exhibit pseudocritical features, such as
the maximum of the heat capacity, whose locations $x_{\rm pc}(L)$
approach the thermodynamic critical point $x_c$.
Their finite-size shift is commonly written
$ x_{\rm pc}(L)=x_c+aL^{-y}$, where $y$ is the shift
exponent~\cite{FerdinandFisher1969,Binder1981}.

\begin{figure}
\centering
\includegraphics[width=\columnwidth]{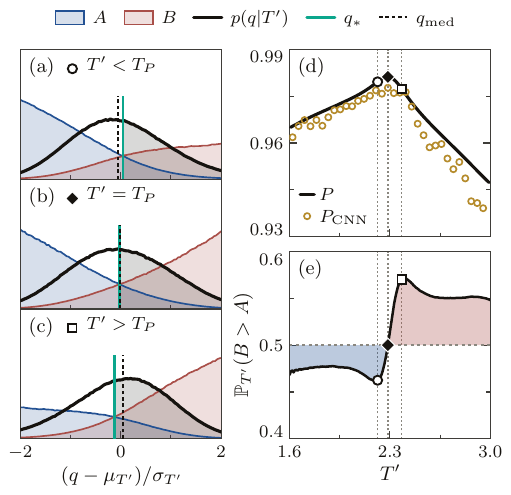}
\caption{
At the confusion peak the classifier assigns half of the samples taken
at the trial point to each class (Ising, $L=32$, grid uniform in $T$,
128\,000 energies per grid temperature).
(a)--(c) Mixtures $A$ and $B$, the distribution $p(q|T')$ with its part
assigned to $B$ shaded, the decision boundary $q_*$, and the median
$q_{\mathrm{med}}$ at $T'$ below, at, and above the confusion peak $T_P$
(circle, diamond, square); energies in standard deviations $\sigma_{T'}$
from the mean $\mu_{T'}$.
(d) Accuracy $P(T')$ of the Bayes-optimal classifier (line) and
$P_{\rm CNN}(T')$ of a trained convolutional network (circles, four
trainings), both on the network's 160 grid temperatures.
(e) Share $\Prob_{T'}(B>A)$ of the samples at $T'$ assigned to $B$
[Eq.~\eqref{eq:balance}].}
\label{fig:balance}
\end{figure}

Machine learning methods offer a new avenue for locating phase
transitions using classifiers trained on raw configurations without
explicit identification and use of an order
parameter~\cite{Carrasquilla2017,HuSinghScalettar2017,Broecker2017,%
Chng2018,Huembeli2018,Venderley2018,Rem2019,Issa2025,Jordan2026}.
For instance, in the learning-by-confusion (LBC)
approach~\cite{vanNieuwenburg2017,Lee2019,Gavreev2022}, a classifier is
trained to distinguish configurations sampled along a sweep of a
parameter $x$ and divided at a proposed critical point $\xp$, the trial
point, into two classes.
When $\xp$ sits at the transition, the two classes correctly represent
the two phases, and the classifier is most highly accurate.
Otherwise, the classifier is less accurate (confused).
The accuracy $P(\xp)$ is W shaped: the middle confusion peak $\xP$
[Fig.~\ref{fig:balance}(d)] is read as the transition point, while the
large values of $P$ at the boundaries of $\xp$ reflect the trivial
achievement of high accuracy when all samples share a single, common
label.
Since its introduction, LBC has been applied to thermal and quantum
critical points, many-body localization, topological transitions,
frustrated magnets, driven-dissipative systems, first-order transitions,
entanglement breakdown, and interacting fermion–boson
models~\cite{vanNieuwenburg2017,Doggen2018MBL,Lee2019,%
Zvyagintseva2022Polariton,Gavreev2022,RichterLaskowska2023Confusion}.

Previous work~\cite{Arnold2022,Arnold2023} established two important
ingredients for understanding finite size effects in LBC: the
Bayes-optimal classifier, and a local connection between classification
performance and Fisher information.
What remained unknown was what fixes the maximum of the standard, global
confusion curve, in which each class contains all sampled parameters on
one side of a trial boundary.
Here we answer that question.
We derive an exact balance condition for the confusion maximum and show
that, to leading order, it places the peak at the Fisher-information
maximum of the parameter used to sample the data.
This has an unexpected consequence: identical physical configurations
can yield different finite-size pseudocritical trajectories when the
same sweep is parametrized uniformly in $T$, ${\rm ln} T$, or $\beta$.
Thus the finite-size output of learning by confusion reflects not only
the underlying physics but also the geometry of the sampling protocol.
Parametrization dependence of LBC was seen empirically:
Ref.~\cite{Arnold2022} found qualitatively different confusion
indicators for Wegner's Ising gauge theory when temperature $T$ and
inverse temperature $\beta$ were used as the tuning parameter.
Here we analyze this parametrization dependence in terms of Fisher
information and finite-size scaling.

In this Letter we first show that the confusion peak tracks the Fisher
information $F_x$ of the sampled parameter.
At the peak, the Bayes-optimal classifier assigns half of the samples
taken at the trial point to each class.
Its Bayes error is $F_x^{-1/2}$, the local parameter scale over which a
single sample becomes sensitive to changes in $x$, times an overlap $c$
obeying $c=\sqrt{2/\pi}$ for Gaussian distributions.
The peak lies at the Fisher maximum wherever $c$ is stationary, moves
away from it otherwise, and inherits its finite-size law.
A second key conclusion concerns the parametrization dependence of the
finite size corrections to the position of the central LBC peak.
We test these results on the Ising, Blume--Capel, three-state Potts, and
Baxter--Wu models.

\textit{Setup.---} We consider a one-parameter equilibrium family whose
probability distribution can be written
\begin{equation}
 p(q|x)=\Omega(q)\exp[-\lambda(x)q-\psi(\lambda(x))],
 \label{eq:expfamily}
\end{equation}
where $x$ is the coordinate in which the data are sampled and $q$ is the
statistic conjugate to the thermodynamic field $\lambda$.
For a temperature sweep, $x$ may be $T$, $\ln T$, or $\beta=1/T$, with
$\lambda=\beta$ and $q$ the energy.
For a Blume--Capel crystal-field sweep at fixed $T$, $\lambda=\Delta/T$
and $q=\sum_{\bf r} S_{\bf r}^2$, the number of occupied sites.
Thus $q$ contains all information about the swept parameter needed by
the Bayes-optimal classifier~\cite{Arnold2022,Arnold2023}.

We sample uniformly in $x$ over $[x_-,x_+]$.
A proposed split $\xp$ assigns samples with $x<\xp$ to class $A$ and
those with $x>\xp$ to class $B$.
The corresponding class mixtures are
\begin{equation}
 A(q;\xp)=\int_{x_-}^{\xp}p(q|x)\dd x,\qquad
 B(q;\xp)=\int_{\xp}^{x_+}p(q|x)\dd x,
 \label{eq:mixtures}
\end{equation}
shown in Figs.~\ref{fig:balance}(a)--\ref{fig:balance}(c).
When the context is clear we suppress the explicit $\xp$ argument.

At fixed $q$, the Bayes-optimal classifier chooses whichever of $A$ or
$B$ is larger~\cite{Arnold2022}.
Its error is
\begin{equation}
 E(\xp)=\int\min(A,B)\dd q,
 \label{eq:error}
\end{equation}
and its accuracy is $P(\xp)=1-E(\xp)/(x_{+}-x_{-})$.
This Bayes accuracy is an upper bound on the accuracy of any classifier,
trained or otherwise~\cite{Arnold2022}.
Although the convolutional network acts on the full spin configuration,
$q$ is a sufficient statistic for the one-parameter equilibrium families
considered here, so the Bayes-optimal classifier of configurations is
equivalent to the Bayes-optimal classifier of $q$.
A trained convolutional network comes within $0.005$ of this bound at
the peak and peaks at the same trial point [Fig.~\ref{fig:balance}(d)].
Without training a neural network, we obtain $E$ from Monte Carlo
histograms of $q$ followed by multiple-histogram
reweighting~\cite{FerrenbergSwendsen1989}.

Throughout this work we distinguish three locations for the phase
transition: $x_c$ is the thermodynamic critical point, $\xP(L)$ the
interior maximum of the LBC accuracy, and $\xF(L)$ the maximum of the
Fisher information in the sampled coordinate.

\textit{Models.---} We test the theory on four classical lattice models
chosen to span distinct types of critical behavior: the two-dimensional
Ising, Blume--Capel, three-state Potts, and Baxter--Wu
models~\cite{Onsager,Ising_review,EIGP2016,blume1966,capel1966,%
Potts_1952,wu1982,wood1972,baxter1973}.
Together they provide ordinary continuous transitions, a tricritical
point, and a first-order transition, as well as different shift
exponents with which to test the predicted finite-size scaling.
Hamiltonians and reference critical parameters are given in the
Supplemental Material.
We generate equilibrium configurations using single-spin Metropolis (Baxter--Wu),
or by Wolff cluster updates combined with Metropolis (Ising, Potts) or heat-bath 
(Blume--Capel) sweeps,
and construct temperature grids uniform in $T$, $\ln T$, and
$\beta$; for the Blume--Capel model we additionally perform
crystal-field sweeps at fixed $T=1.2$, $0.608$, and $0.6$.

\textit{The balance condition behind the confusion peak.---} The
location of the confusion peak follows from a simple exact balance
condition.
Increasing the trial split by $d\xp$ transfers the samples at $\xp$ from
class $B$ to class $A$.
Equivalently, the two class mixtures change according to
$ \partial_{x'}A(q;x')=p(q|x'),\qquad \partial_{x'}B(q;x')=-p(q|x'), $
so moving the trial point simply transfers an infinitesimal slice of
probability from class $B$ to class $A$.
Samples at $\xp$ which the Bayes classifier assigns to $B$ therefore
become misclassified, while those assigned to $A$ become correctly
classified.
Consequently,
\begin{equation}
 \frac{\dd E}{\dd\xp} = \Prob_{\xp}(B>A)-\Prob_{\xp}(A>B).
 \label{eq:balance}
\end{equation}
Equation~\eqref{eq:balance} is exact for any choice of classifier input
and any sampling grid, provided the Bayes-optimal decision rule is used.
Assuming ties carry no probability, the confusion curve is stationary
exactly when the classifier assigns equal probability from the
distribution at the trial point to the two classes.
For the Ising example, $\Prob_{\xp}(B>A)$ is $0.46$, $0.50$, and $0.57$
below, at, and above the peak [Fig.~\ref{fig:balance}(e)].

For the Boltzmann distributions considered here, which have the
exponential-family form of Eq.~\eqref{eq:expfamily}, $\Omega(q)$ cancels
from $A/B$, which is monotonic in $q$~\cite{LehmannRomano2022}.
The two mixtures therefore cross once at a decision boundary $\qs$.
Equation~\eqref{eq:balance} then becomes the transparent condition
\begin{equation}
 \qs(\xP)=\operatorname{median}_{p(q|\xP)} q .
 \label{eq:median}
\end{equation}
Thus the global LBC optimum is fixed by a boundary that depends on the
entire sweep, but which at the peak bisects the distribution at the
trial point [Figs.~\ref{fig:balance}(a)--\ref{fig:balance}(c)].

\textit{The Fisher scale.---} The samples hardest to classify lie near
the trial point, where neighboring equilibrium distributions strongly
overlap.
The natural scale governing how rapidly $p(q|x)$ changes is the Fisher
information in the sampled coordinate,
\begin{equation}
 F_x(x)=\big\langle[\partial_x\ln p(q|x)]^2\big\rangle_x .
 \label{eq:fisher}
\end{equation}
Writing $g=|dx/d\lambda|$ and denoting the variance of $q$ by
$\sigma^2$, Eq.~\eqref{eq:expfamily} gives
\begin{equation}
 F_x=\frac{\sigma^2}{g^2}.
 \label{eq:fishermetric}
\end{equation}
Hence $F_x^{-1/2}=g/\sigma$ is the local parameter interval required to
move the distribution by roughly one standard deviation.
For thermal sweeps, with $C=\sigma^2/T^2$ the heat capacity,
\begin{equation}
 F_T=\frac{C}{T^2},\qquad F_{\ln T}=C,\qquad F_\beta=CT^2,
 \label{eq:threefisher}
\end{equation}
so the same configurations can have different Fisher maxima on grids
uniform in $T$, $\ln T$, and $\beta$.

The Bayes error can be factored as
\begin{equation}
 E(x)=\frac{c(x)}{\sqrt{F_x(x)}},
 \label{eq:factor}
\end{equation}
where the dimensionless factor $c(x)\equiv E\sqrt{F_x}$ contains the
residual dependence on the shapes of the distributions and on the finite
sweep.
Therefore,
\begin{equation}
 \partial_x\ln E=\partial_x\ln c-\tfrac12\,\partial_x\ln F_x .
 \label{eq:tilt}
\end{equation}
If $c$ is stationary at the Fisher maximum, then $\xP=\xF$ exactly; if
$c$ varies slowly there, the displacement is small.
For a Gaussian reference family one obtains
\begin{equation}
 E_0=\sqrt{\frac2\pi}\,\frac1{\sqrt{F_x}},
 \label{eq:gaussian}
\end{equation}
so $c_0=\sqrt{2/\pi}\simeq0.80$.
The derivation of Eq.~\eqref{eq:gaussian}, the symmetry conditions for
exact peak coincidence, and the first-order two-phase limit are given in
the Supplemental Material.
\begin{figure}
\centering
\includegraphics[width=\columnwidth]{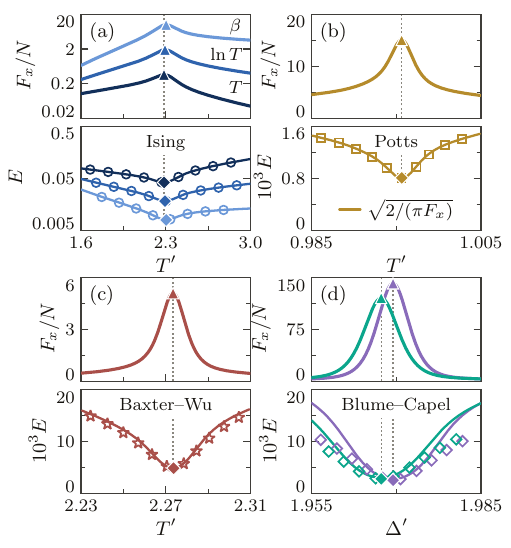}
\caption{
The error follows the Fisher information.
Upper panels: $F_x/N$ [Eq.~\eqref{eq:fisher}], $N$ the number of sites,
with its maximum $\xF$ (triangle, dotted line); lower panels:
$E=(x_{+}-\,x_{-})\,(1-P)$ (hollow markers), the Gaussian curve
$\sqrt{2/(\pi F_x)}$ of Eq.~\eqref{eq:gaussian} (line), and the
confusion peak $\xP$ (diamond).
(a) Ising ($L=32$) on grids uniform in $T$, $\ln T$, and $\beta$ (dark
to light; log axes, each in its grid's variable); (b) three-state Potts
($L=256$) and (c) Baxter--Wu ($L=72$) on grids uniform in $T$; (d)
Blume--Capel $\Delta$ sweeps through the tricritical point ($T=0.608$,
teal) and the first-order line ($T=0.6$, violet), $L=24$.
The number of samples per grid point was 128\,000 (a, b), 8000 (c), and
1200 (d).}
\label{fig:fisher}
\end{figure}

\begin{figure}
\centering
\includegraphics[width=\columnwidth]{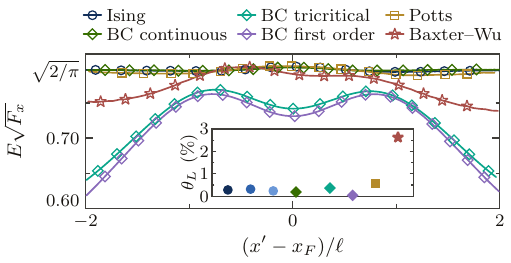}
\caption{
The overlap sets the height of the error, and its slope the position of
the peak: $c=E\sqrt{F_x}$ against $(\xp-\xF)/\ell$ for Ising ($T$ grid,
$L=64$), the Blume--Capel (BC) $\Delta$ sweeps at $T=1.2$ (continuous,
$L=64$), $0.608$ (tricritical, $L=48$), and $0.6$ (first order,
$L=24$), Potts ($L=256$), and Baxter--Wu ($L=72$).
Inset: $\theta_L$ [Eq.~\eqref{eq:theta}] in the order of the key, Ising
on its $T$, $\ln T$, and $\beta$ grids (dark to light), averaged over
$L=24$--$64$, $24$--$64$, $24$--$48$, $16$--$24$, $96$--$256$, and
$36$--$72$; error bars mostly within the markers.}
\label{fig:offset}
\end{figure}

\vskip0.10in
\textit{Peak coincidence and separation.---} Equation~\eqref{eq:tilt}
makes the role of the overlap factor explicit.
A reflection-symmetric sweep forces $E$, $F_x$, and $c$ to be stationary
at the symmetry point, so if that point is the Fisher maximum the LBC
and Fisher peaks coincide independently of the detailed shape of the
distributions.
More generally, Fig.~\ref{fig:offset} shows that $c$ varies only weakly
through $\xF$ across all of the models studied here.
The Gaussian value $\sqrt{2/\pi}$ describes the continuous Ising,
Blume--Capel, and Potts sweeps particularly well, while tricritical and
first-order coexistence lowers the overlap without substantially moving
the peak.
The detailed reflection argument, including which non-Gaussian
corrections can shift the peak, is given in the Supplemental Material.

We measure the residual displacement in units of the Fisher-peak width
$\ell$, defined locally by $\ell^2=-2/[\partial_x^2\ln F_x]_{\xF}$,
\begin{equation}
 \theta_L=\frac{\xP-\xF}{\ell}.
 \label{eq:theta}
\end{equation}
Across the six sweeps in Fig.~\ref{fig:offset}, the two peaks differ by
at most $2.62(6)\,\%$ of $\ell$, with the largest displacement in the
Baxter--Wu model.
Thus the Fisher maximum controls the location of the global LBC peak
even when the underlying distributions are strongly non-Gaussian.

\begin{figure}
\centering
\includegraphics[width=\columnwidth]{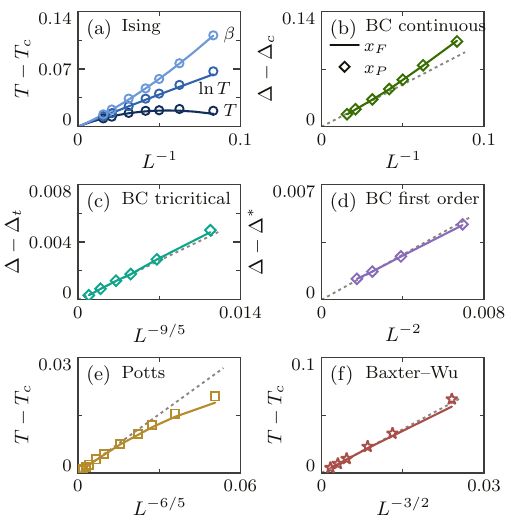}
\caption{
The confusion peak inherits the finite-size law of the Fisher maximum:
distance of $\xF$ (lines) and $\xP$ (hollow markers) from the critical
point against $L^{-y}$.
(a) Ising on its $T$, $\ln T$, and $\beta$ grids (dark to light), with
the exact torus maxima~\cite{Kaufman1949,Beale1996} up to $L=1024$;
(b)--(d) Blume--Capel (BC) $\Delta$ sweeps at $T=1.2$
($\Delta_c=1.4167$, $y=1$), $0.608$ ($\Delta_t=1.9660$, $y=9/5$), and
$0.6$ ($\Delta^*=1.968174$, $y=2$); (e) Potts ($y=6/5$);
(f) Baxter--Wu ($y=3/2$).
Dotted, (b)--(f): $a_FL^{-y}$ fitted at the largest sizes.
Error bars are smaller than the markers.}
\label{fig:scaling}
\end{figure}

\textit{Finite-size scaling.---}The Fisher scale also fixes how the
confusion peak moves with $L$.
Since $\theta_L$ measures the displacement in widths of the Fisher peak,
$\xP$ inherits the finite-size law of $\xF$ whenever that width shrinks
as fast as the shift.
Finite-size scaling of the free energy gives each $\lambda$ derivative
of $\psi$ a factor $L^y$, with
$y=1/\nu$~\cite{FerdinandFisher1969,Binder1981}.
If $2y>d$, with $d$ the physical dimension, the singular part of
$\sigma^2=\partial_\lambda^2\psi$ outgrows its regular part, which is
proportional to the number of sites $L^d$.
Consequently, the Fisher peak narrows as $\ell_L=\ell_0L^{-y}$ and its
maximum shifts as $\xF=x_c+a_FL^{-y}$, like any pseudocritical point.
If we further assume that the normalized error curve, the error measured
from $\xF$ in widths $\ell_L$ and divided by $E(\xF)$, converges to a
limiting shape with a single minimum $\theta_\infty$, then
$\theta_L\to\theta_\infty$, and since $\xP=\xF+\ell_L\theta_L$,
\begin{equation}
 \xP(L)=x_c+(a_F+\theta_\infty\ell_0)L^{-y}+o(L^{-y}).
 \label{eq:fss}
\end{equation}
The confusion peak has the exponent of the Fisher maximum of its grid,
and the displacement only changes the amplitude.

The Blume--Capel tricritical ($y=9/5$), Potts, and Baxter--Wu sweeps
satisfy $2y>d$, and in each $\xP$ follows $\xF$ from size to size
[Figs.~\ref{fig:scaling}(c), \ref{fig:scaling}(e), and
\ref{fig:scaling}(f)].
Baxter--Wu, whose displacement is the largest, shows the change of
amplitude: at $L=36$--$72$, $(\xP-x_c)L^{3/2}=2.65$--$2.73$ exceeds
$(\xF-x_c)L^{3/2}=2.52$--$2.61$ by $\theta_\infty\ell_0\approx0.13$
[Fig.~\ref{fig:scaling}(f)].

Two cases fall outside the derivation.
In the Ising class, $2y=d$.
The heat capacity grows only as $\ln L$~\cite{FerdinandFisher1969}, and
the width shrinks as $\sqrt{\ln L}/L$, more slowly than the shift.
On the Blume--Capel first-order line, $y=d=2$ reflects the coexistence
of two phases, not a diverging correlation length.
However, in both cases, $\xP$ stays on $\xF$ at every size
[Figs.~\ref{fig:scaling}(a), \ref{fig:scaling}(b), and
\ref{fig:scaling}(d)].

The confusion peak thus follows the Fisher maximum of the variable in
which the sweep is uniform.
The Ising model sampled on grids uniform in $T$, $\ln T$, and $\beta$
gives three different shifts, a factor of two apart at $L=32$, with
$\xP$ on each [Fig.~\ref{fig:scaling}(a)].
On the $\ln T$ grid $\xF$ is the familiar heat-capacity
maximum~\cite{FerdinandFisher1969}.
On the $T$ grid the extra factor $1/T^2$ pulls $\xF$ toward lower $T$ by
an amount proportional to $\ell^2$, which at small $L$ removes much of
the shift: $T_F-T_c$ stays at $0.016$--$0.019$ for $L=12$--$32$.
This is why Ising confusion peaks on temperature grids barely move with
size.
Since $\ell^2$ falls faster than the shift, the grid changes only the
corrections to Eq.~\eqref{eq:fss}, and the three shifts approach a
common law [Fig.~\ref{fig:scaling}(a)].

\vskip0.10in
\textit{Conclusions.---} In summary, we have identified what the central
peak of learning by confusion measures.
The Bayes-optimal peak obeys an exact balance condition, and its natural
scale is set by the Fisher information of the parameter used to generate
the sweep.
When the dimensionless overlap of the two classes varies slowly, the
confusion peak tracks the Fisher maximum and therefore inherits its
finite-size scaling.
This explains, within a single framework, its behavior at continuous,
tricritical, and first-order transitions.

An important consequence is that a finite-size machine-learning estimate
need not be a property of the physical system alone.
For the same equilibrium configurations, sampling uniformly in $T$,
${\rm ln} T$, or $\beta$ changes the relevant Fisher information and
hence changes the finite-size trajectory of the confusion peak.
The specification of the sampling coordinate should therefore be
regarded as part of the definition of an LBC estimate.

Our exact balance condition applies to the original LBC formulation,
in which the class priors vary with the position of the trial split.
A balanced variant with equal class priors has recently been proposed
to remove this class-imbalance contribution~\cite{arnold2024}.
Whether its global optimum obeys an analogous relation to Fisher
information and exhibits the same finite-size scaling is an interesting
open question.

Although our exact results apply to learning by confusion, the
underlying issue is more general.
Many machine-learning approaches to phase detection ultimately
distinguish probability distributions generated at nearby points in
parameter space.
Fisher information provides the natural local metric for that
distinguishability.
It will therefore be important to determine, for other supervised and
unsupervised approaches, which statistical distance or information
metric their outputs probe and whether their finite-size estimates
likewise depend on the parametrization and sampling measure.
In this sense, the present work provides both an explanation of learning
by confusion and a broader caution: before interpreting an ML-derived
pseudocritical point as a property of the physics, one should understand
which features of the data-generation protocol the algorithm is also
measuring.

\vskip0.10in
\textit{Acknowledgments:}
GI and RTS were both supported by the grant DOE DE-SC0014671 funded by
the U.S.~Department of Energy, Office of Science.

\nocite{apsrev42Control}
\bibliographystyle{apsrev4-2}
\bibliography{letter}

\clearpage
\setcounter{equation}{0}
\renewcommand{\theequation}{SM.\arabic{equation}}

{\bf Supplemental Material.}
In these Supplemental Materials we provide (1) definitions of the
classical statistical-mechanics models, (2) the Gaussian reference
calculation and reflection argument underlying the near-coincidence of
the LBC and Fisher maxima, and (3) the first-order two-phase limit.

\vskip0.10in \noindent
{\bf 1.~Models}

The Ising model~\cite{Onsager,Ising_review,EIGP2016}
\begin{equation}
H = - J \sum_{\langle {\bf r r'} \rangle} S_{\bf r} S_{\bf r'}
 \label{eq:ising}
\end{equation}
describes classical `spin' variables $S_{\bf r} = \pm 1$ located on the
sites ${\bf r}$ of a spatial lattice and interacting in near neighbor
pairs $\langle {\bf r r'} \rangle$ via an (exchange) energy $J$.
Here we consider a square geometry.
In this case, for $J>0$, the model undergoes a second order paramagnetic
(PM) to ferromagnetic (FM) transition at
$T_c/J= 2/{\rm ln}(1+\sqrt{2}) \sim 2.269$.
Here we have used learning by confusion to determine this $T_c$ by
splitting Monte Carlo data with PM and FM labels based on a guessed
$T^{\prime}$.

The Blume-Capel model~\cite{blume1966,capel1966}
\begin{equation}
H = - J \sum_{\langle {\bf r r'} \rangle} S_{\bf r} S_{\bf r'} + \Delta \sum_{\bf r} S_{\bf r}^2
 \label{eq:blumecapel}
\end{equation}
generalizes the Ising model to allow for vacancies, $S_{\bf r} = 0$ in
addition to occupied sites $S_{\bf r}=\pm 1$.
A crystal field energy $\Delta < 0 \,\,(> 0)$ favors(penalizes)
$S_{\bf r}=\pm1$.
The Ising model is recovered in the limit $\Delta \rightarrow -\infty$.
The critical temperature decreases as $\Delta$ increases from this
limit, ultimately vanishing at $\Delta/J=2$ for the square lattice.
The transition also changes from second to first order at the
tricritical point $(T_t/J,\Delta_t/J) = (0.608, 1.966)$.
We make use of the expanded parameter space to investigate splits
involving the crystal field energy $\Delta$.

The Potts model~\cite{Potts_1952,wu1982}
\begin{equation}
H = - J \sum_{\langle {\bf r r'}\rangle} \delta_{S_{\bf r},S_{\bf r'}}
 \label{eq:potts}
\end{equation}
has $S_{\bf r}=1,2,\cdots Q$ taking $Q$ possible values.
The $Q=2$ case is equivalent to the Ising model up to a rescaling of $J$
by a factor of $2$.
When $ Q \leq 4$ the transition to FM is second order, but for $Q>4$
the transition becomes first order.
On a square lattice, the critical temperature is
$T_c(Q)/J= 1/{\rm ln}(1+\sqrt{Q})$.

The Baxter-Wu model~\cite{wood1972,baxter1973}
\begin{equation}
H = - J \sum_{\langle {\bf r r' r''} \rangle} S_{\bf r} S_{\bf r'} S_{\bf r''}
 \label{eq:baxterwu}
\end{equation}
involves three spin interactions on each of the plaquettes
$\langle {\bf r r' r''} \rangle$ of a triangular lattice.
Although the Baxter-Wu model lacks the inversion symmetry
$S_{\bf r} \leftrightarrow -S_{\bf r}$ of Ising and Blume-Capel, it has
its own discrete symmetry structure leading to four degenerate ground
states.
The critical temperature is identical to that of the Ising model,
however the universality class is that of the $Q=4$ Potts model.

\vskip0.10in \noindent
{\bf 2.~Why the confusion and Fisher maxima nearly coincide}

The main text writes the Bayes error as
\begin{equation}
 E(x)=\frac{c(x)}{\sqrt{F_x(x)}} ,
 \label{eq:sm_factor}
\end{equation}
where $F_x$ is the Fisher information in the sampled coordinate and
$c(x)$ is a dimensionless overlap factor.
We give here several useful limits that explain why $c(x)$ is often
nearly stationary at the Fisher maximum.

\textit{Gaussian reference problem.---} Consider a family in which the
sufficient statistic $q$ is Gaussian with constant standard deviation
$\sigma$, while its mean $\mu(x)$ moves at a constant rate
$\mu'=d\mu/dx$.
For a sweep broad enough that the boundaries can be neglected near the
trial point $\xp$, define $r=[q-\mu(\xp)]/\sigma$.
The two class mixtures are proportional to
\begin{equation}
 A(r)=\Phi(-r),\qquad B(r)=\Phi(r),
\end{equation}
where $\Phi$ is the standard normal cumulative distribution function.
The decision boundary is $r=0$, i.e.~the mean and median of the split
distribution.
Since $F_x=\mu'^{\,2}/\sigma^2$, direct integration gives
\begin{equation}
 E_0=\sqrt{\frac{2}{\pi}}\frac{1}{\sqrt{F_x}},\qquad
 c_0=\sqrt{\frac{2}{\pi}}\simeq0.798 .
 \label{eq:sm_gaussian}
\end{equation}
Thus the Bayes error is, up to a dimensionless factor of order unity,
the parameter interval required to shift the mean of the sufficient
statistic by one standard deviation.

\textit{Reflection symmetry.---} The coincidence of the confusion and
Fisher maxima does not require Gaussian distributions.
Suppose a sweep is its own mirror image about $x_0$, so that for some
$q_0$
\begin{equation}
 p(q|x_0+v)=p(2q_0-q|x_0-v)
 \label{eq:sm_reflection}
\end{equation}
throughout the relevant part of the sweep.
Reflection exchanges the two global classes at trial points $x_0+v$ and
$x_0-v$.
Consequently $E$, $F_x$, and $c$ are even about $x_0$ and therefore
stationary there.
If $x_0$ is the Fisher maximum and the error minimum, then $x_P=x_F=x_0$
exactly, independently of the detailed distribution shape.

This argument also shows which deviations can move the confusion peak.
Departures unchanged under reflection are even; examples include changes
in kurtosis.
Departures that reverse sign are odd; examples include skewness, a slope
of the sampling density, and unequal distances to the two ends of the
sweep.
Even departures can alter the height of the error without shifting its
minimum.
Odd departures are required for a displacement.
The leading odd corrections shift both the decision boundary $q_*$ and
the median of $p(q|\xp)$~\cite{FisherCornish1960}; to first order their
difference is proportional to $\partial_x\ln F_x$.
Hence the exact condition $q_*=\operatorname{median}q$ is, to this
order, also the Fisher condition $\partial_xF_x=0$.
Only higher-order odd terms, or products of odd and even corrections,
generate a nonzero slope of $c$ at $x_F$ and separate the two peaks.

This explains the Ising results on grids uniform in $T$, $\ln T$, and
$\beta$.
Although the skewness and sampling metric differ among the three
parametrizations, their peak displacements remain at the same
sub-percent level when measured in units of the Fisher-peak width.

\vskip0.10in \noindent
{\bf 3.~First-order two-phase limit}

A strongly first-order transition provides an opposite, maximally
non-Gaussian limit.
Consider approximating the sufficient statistic by two sharp values
$q=\pm a$ representing the two coexisting phases.
At their equal-weight point $\lambda_0$ on a grid uniform in the natural
field $\lambda$, the sweep is reflection symmetric and the confusion and
Fisher extrema coincide.
Direct evaluation gives
\begin{equation}
 E(\lambda_0)=\frac{\ln2}{\sqrt{F_\lambda(\lambda_0)}},
 \qquad c(\lambda_0)=\ln2\simeq0.693 .
 \label{eq:sm_twophase}
\end{equation}
Thus coexistence changes the height of the confusion error substantially
relative to the Gaussian value $\sqrt{2/\pi}$ without moving the
location of its minimum.
This explains why the first-order Blume--Capel data can lie visibly
below the Gaussian reference curve while the LBC and Fisher maxima
remain coincident to high accuracy.

Away from exact symmetry, Eq.~\eqref{eq:tilt} shows that the separation
is controlled by the slope of $c$ at the Fisher maximum.
The small measured slopes in Fig.~\ref{fig:offset} therefore account
directly for the small values of $\theta_L$ reported in the main text.

\end{document}